\documentclass[sigconf]{acmart}

\AtBeginDocument{%
  }

\setcopyright{acmlicensed}
\copyrightyear{2018}
\acmYear{2018}
\acmDOI{XXXXXXX.XXXXXXX}
\acmConference[Conference acronym 'XX]{Make sure to enter the correct
  conference title from your rights confirmation email}{June 03--05,
  2018}{Woodstock, NY}

\usepackage{booktabs}
\usepackage{multirow}
\usepackage{tabularx}
\usepackage{cleveref}
\usepackage{hyperref}
\hypersetup{colorlinks=true,citecolor=brown}
\usepackage{url}
\usepackage[table,xcdraw,dvipsnames]{xcolor}
\usepackage[most]{tcolorbox}
\usepackage{graphicx}

\usetikzlibrary{shapes.geometric, arrows.meta, positioning, calc, backgrounds}

\definecolor{scoregreen}{RGB}{34, 139, 34}

\newtcolorbox{PromptBox}[1][]{
  enhanced, breakable,
  colback=gray!3,
  colframe=gray!70!black,
  boxrule=0.5pt,
  arc=2mm,
  left=2mm,right=2mm,top=2mm,bottom=2mm,
  fonttitle=\bfseries\small,
  coltitle=white,
  colbacktitle=black!85,
  attach boxed title to top left={xshift=2mm,yshift=-2mm},
  boxed title style={size=small, frame hidden, interior style={fill=black!85}},
  #1
}

\usepackage{xcolor}
\usepackage{soul}

\definecolor{cb-vermillion}{HTML}{D55E00} 
\definecolor{cb-skyblue}{HTML}{56B4E9}    
\definecolor{cb-amber}{HTML}{F5C710}      

\newtcolorbox{InstructionBox}[1][]{
  enhanced, breakable,
  colback=gray!5,
  colframe=gray!60,
  boxrule=0.4pt,
  arc=1.5mm,
  left=2mm,right=2mm,top=1.5mm,bottom=1.5mm,
  #1
}

\newtcolorbox{FigureInstructionBox}[1][]{
  enhanced,
  colback=gray!5,
  colframe=gray!60,
  boxrule=0.4pt,
  arc=1.5mm,
  left=2mm,right=2mm,top=1.5mm,bottom=1.5mm,
  #1
}

\usepackage{tabularx}
\usepackage{array}
\usepackage{booktabs}
\usepackage{capt-of} 

\makeatletter
\let\@authornote\@gobble
\def\authornote#1{%
  \appto{\@authornotes}{%
    \footnotetext{#1}%
  }%
}
\makeatother

\begin{document}

\title{Deploying Semantic ID-based Generative Retrieval for Large-Scale Podcast Discovery at Spotify}

\author{
    Edoardo D'Amico, Marco De Nadai, Praveen Chandar, Divita Vohra, Shawn Lin, 
    Max Lefarov\texorpdfstring{\textsuperscript{$\dagger$}}{1}, Paul Gigioli, Gustavo Penha, Ilya Kopysitsky, Ivo Joel Senese, 
    Darren Mei, Francesco Fabbri, Oguz Semerci, Yu Zhao\texorpdfstring{\textsuperscript{$\dagger$}}{2}, Vincent Tang, Brian St. Thomas, Alexandra Ranieri, Matthew N.K. Smith, Aaron Bernkopf, Bryan Leung, Ghazal Fazelnia, Mark VanMiddlesworth, Timothy Christopher Heath, Petter Pehrson Skiden, Alice Y. Wang, Doug J. Cole, Andreas Damianou, Maya Hristakeva, Reid Wilbur, Tarun Chillara, Vladan Radosavljevic, Pooja Chitkara, Sainath Adapa, Juan Elenter, Bernd Huber, Jacqueline Wood, Saaketh Vedantam\texorpdfstring{\textsuperscript{$\dagger$}}{3}, Jan Stypka\texorpdfstring{\textsuperscript{$\dagger$}}{4}, Sandeep Ghael, Martin D. Gould, David Murgatroyd, Yves Raimond, Mounia Lalmas, Paul N. Bennett
}

\affiliation{%
  \institution{Spotify}
  \country{Spain, Denmark, United States, Germany, France, United Kingdom}
}

\email{edoardod,mdenadai,praveenr,divitav,weihsiangl,pgigioli,gustavop,ikopysitsky,ivojoels,darrenm,francescof,oguz, vtang, brianstt,}
\email{
asimonoff,mattnks,abernkopf,bleung,ghazalf,mvanmiddlesworth,theath,peppe,alicew,dougcole,andreasd,mayah,reid,tarunc,}
\email{vladanr,poojac,sainatha,juane,berndh,jacquelinew,sandeepg,mg,dmurga,yvesr,mounial,pbennett@spotify.com}

\authornote{$\dagger$ Work performed while at Spotify.}

\renewcommand{\shortauthors}{Edoardo D'Amico et al.}

\begin{abstract}

Podcast listening is often grounded in a set of favorite shows, while listener intent can evolve over time. This combination of stable preferences and changing intent motivates recommendation approaches that support both familiarity and exploration. Traditional recommender systems typically emphasize long-term interaction patterns, and are less explicitly designed to incorporate rich contextual signals or flexible, intent-aware discovery objectives. In this setting, models that can jointly reason over semantics, context, and user state offer a promising direction. Large Language Models (LLMs) provide strong semantic reasoning and contextual conditioning for discovery-oriented recommendation, but deploying them in production introduces challenges in catalog grounding, user-level personalization, and latency-critical serving.

We address these challenges with GLIDE, a production-scale generative recommender for podcast discovery at Spotify. GLIDE formulates recommendation as an instruction-following task over a discretized catalog using Semantic IDs, enabling grounded generation over a large inventory. The model conditions on recent listening history and lightweight user context, while injecting long-term user embeddings as soft prompts to capture stable preferences under strict inference constraints. We evaluate GLIDE using offline retrieval metrics, human judgments, and LLM-based evaluation, and validate its impact through large-scale online A/B testing. Across experiments involving millions of users, GLIDE increases non-habitual podcast streaming on Spotify home surface by up to 5.4\% and new-show discovery by up to 14.3\%, while meeting production cost and latency constraints.

\end{abstract}


\maketitle

\section{Introduction}

Podcast consumption often exhibits continuity, with listeners regularly engaging with shows\footnote{We use \emph{show} to denote a podcast series and \emph{episode} to denote an individual installment within a show.} they already follow. Supporting effective discovery alongside this continuity remains an important challenge for recommender systems~\cite{nazari2022recommending}.
When recommendation models rely predominantly on historical interactions, they tend to emphasize familiar content, while adapting recommendations to evolving listener intent and surfacing timely, relevant episodes requires additional modeling capacity.

Listener intent is inherently dynamic: even highly engaged users may seek different content depending on context, such as timely updates during a commute, brief entertainment between meetings, or deeper topical exploration at home. Approaches based on similarity heuristics or static user–item formulations are not explicitly designed to capture such shifts within or across sessions. Addressing this challenge calls for modeling listening behavior as a sequence that integrates short-term context with stable preferences, while preserving a robust semantic understanding of podcast topics.

These requirements motivate Large Language Models (LLMs) as a promising foundation for podcast recommendation. LLMs support long-context sequence modeling, instruction-conditioned objectives, and rich pre-trained semantic representations. However, applying them in recommender systems introduces additional challenges, including grounding model outputs to large, fast-changing catalogs, achieving long-term user-level personalization, and meeting strict latency and reliability constraints. 

We address these challenges by proposing GLIDE  (\textbf{G}rounded \textbf{L}LM for \textbf{I}nterest \textbf{D}iscovery r\textbf{E}commendations), a generative, language-based podcast recommender that frames recommendation as an instruction-following task. Given a  prompt encoding recent listening history, lightweight user context, and an explicit discovery objective, the model generates identifiers for episodes the user is likely to listen next. Episodes are represented using \emph{Semantic IDs (SIDs)}~\cite{rajput2024recommender}, which allow the model to generate valid catalog items directly while providing a compact, semantically meaningful interface to a large and evolving podcast catalog.
To incorporate long-term personalization without inflating prompt length, we inject compact soft prompts derived from collaborative-filtering user embeddings~\cite{huber2025embedding, 10.1145/3705328.3748132}.

We evaluate GLIDE using offline retrieval metrics, human judgments, and LLM-based evaluation, and assess its impact through large-scale online A/B testing on Spotify. Beyond accuracy, we report practical lessons on data construction, model adaptation, and serving under real-world latency and reliability constraints.

\paragraph{Contributions.} 

\begin{itemize}
    \item We introduce an \textbf{instruction-following formulation} for podcast recommendation, enabling session-aware and controllable discovery at the episode level.
    \item We propose a \textbf{practical adaptation recipe} for open-weight LLMs that combines SIDs with compact user-level personalization in a single, promptable recommender architecture.
    \item We present a \textbf{comprehensive evaluation strategy} for discovery-oriented recommendation, combining offline metrics, human judgments, and LLM-based evaluation.
    \item We demonstrate \textbf{production-scale impact} through an online A/B test, showing significant improvements in non-habitual discovery while meeting real-world latency and reliability constraints.
      
\end{itemize}

Together, these contributions demonstrate how language-grounded generative models can be adapted into practical, personalized recommender systems operating at scale.

\section{Related Work}

\paragraph{Podcast Recommendation.}

Episodes are long-form, topical coverage is broad, and listening behavior often exhibits continuity, with users regularly returning to shows they follow. Prior work has studied show-level recommendation~\cite{tsagkias2009podcast, benton2020trajectory}, long-term consumption dynamics~\cite{maystre2023optimizing, mcdonald2023impatient}, and cold-start scenarios, as well as drawing parallels between podcasts and other content modalities~\cite{10.1145/3589335.3648339}. Our work complements these directions by focusing on recommendations that support discovery alongside established listening patterns, a setting that poses distinct modeling challenges for recommender systems~\cite{nazari2022recommending}.

\paragraph{Generative Recommendations.}
Sequential recommenders such as SASRec~\cite{kang2018self} model next-item prediction from interaction histories, typically representing items as atomic identifiers. While effective, atomic ID-based representations often struggle to generalize when interaction data are sparse, particularly for cold-start or long-tail items that lack sufficient historical signals~\cite{singh2024better}. Recent work has introduced SIDs, which discretize continuous content representations into short sequences of tokens, preserving semantic similarity while remaining compatible with sequence models (e.g., TIGER~\cite{rajput2024recommender}) and improving generalization~\cite{singh2024better}. This line of work has sparked broader interest in \emph{grounding} recommendation over tokenized catalogs and in designing \emph{discretization schemes} that balance semantic expressiveness, stability (i.e., maintaining consistent item representations as catalogs and embeddings evolve), and computational efficiency~\cite{tay2022transformer, rajput2024recommender}. Our work builds on this line of research by adopting SIDs for catalog grounding.

\paragraph{LLMs for recommendation and catalog grounding.}
LLMs have recently been adapted to recommendation via text-based prompting and instruction tuning. P5~\cite{geng2022recommendation} introduced a unified, prompt-based framework for multiple recommendation tasks, but relies on atomic item identifiers. Other approaches represent items using textual fields such as titles~\cite{zhang2023recommendation}, which can improve semantic understanding but often incur higher serving cost due to longer prompts and the need to disambiguate noisy  or inconsistent item text at scale. More recent work incorporates SIDs into LLMs to enable catalog-grounded generation and retrieval at scale. For example, PLUM~\cite{2025_PLUMAdaptingPretrainedLanguageModels} adapts pretrained language models into retrieval-style recommenders grounded in a large video catalog, demonstrating the feasibility of SID-based grounding at industrial scale. However, they do not handle textual understanding for discovery tasks, and their deployment setting and related learnings are unknown.

We build on these directions by combining (i) an open-weight LLM trained to operate over catalog-grounded SIDs, (ii) instruction-conditioned generation that captures session-level recommendation objectives, and (iii) soft-prompt personalization via dense collaborative-filtering user embeddings~\cite{huber2025embedding}. While prior work typically focuses on either atomic item identifiers or long textual item representations, we adopt a representation that supports both semantic grounding and efficient generation at scale. In addition, our work targets production deployment over a large, fast-changing podcast catalog, and provides an end-to-end account spanning problem formulation, model adaptation, system design, and online experimentation.


\begin{figure}[t]
  \centering
  \includegraphics[width=\columnwidth]{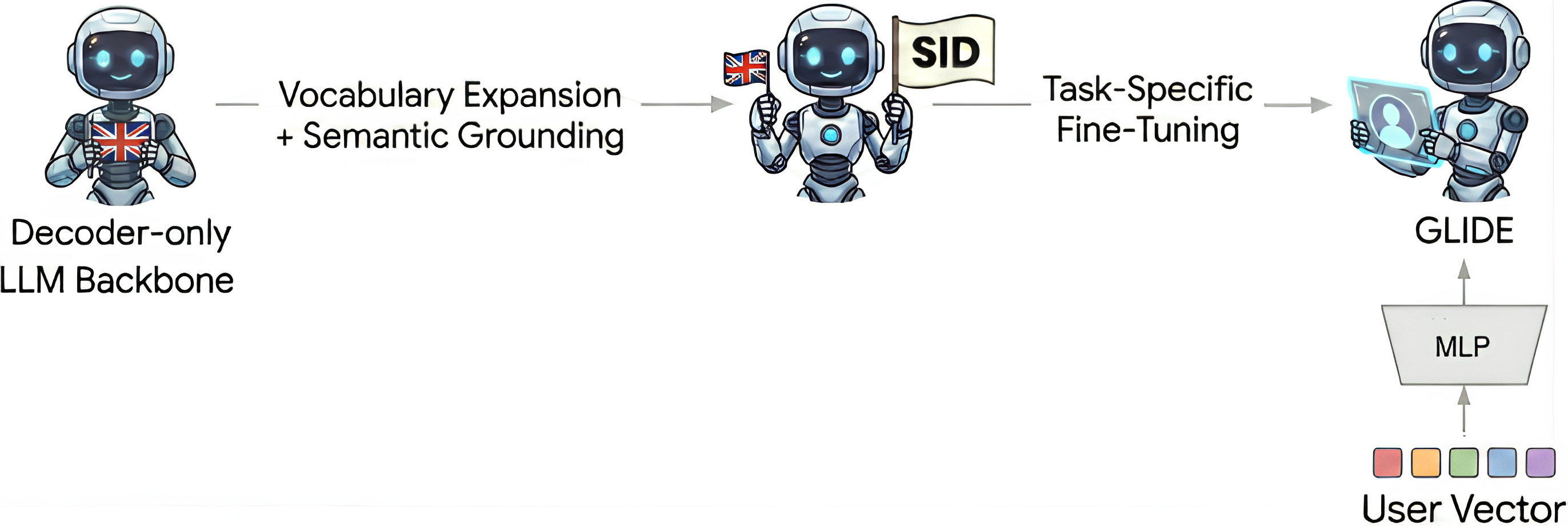}
  \caption{We extend the vocabulary of a pre-trained LLM with SID tokens. Then, we undergo a two phase training. First, we ground the model to \emph{speak} SIDs, along with the original languages. Then, we add a soft prompt user vector projection to add individual-level personalization and train the model for language-steerable podcast recommendations.}
\  \label{fig:architecture}
\end{figure}

\section{Problem design}
\label{sec: problem_design}

We formalize discovery through the notion of \emph{non-habitual} streaming at the user-show level. For a user $u$ and a show $s$, $T_{u,s}$ denotes the total listening time on show $s$ in the past 28 days. We define:
\begin{itemize}
  \item {\bf Habitual shows:} shows with $T_{u,s} \ge 10$ minutes in the past 28 days.
  \item {\bf Non-habitual shows:} shows with $T_{u,s} < 10$ minutes in the past 28 days, which we further subdivide into:
  \begin{enumerate}
    \item \textbf{Non-habitual and unfamiliar:} shows with no historical consumption by $u$ (never listened).
   \item \textbf{Non-habitual but familiar:} shows with some historical consumption, but without a current habit ($T_{u,s} < 10$ minutes in the past 28 days).
  \end{enumerate}
\end{itemize}
Based on empirical analysis of listening behavior and internal user research, the 28-day window captures recent habits while smoothing short-term noise, and the 10-minute threshold distinguishes meaningful engagement from incidental listening.

All other previously consumed shows are treated as habitual for the user. These categories are used throughout to define training targets, evaluation slices, and language-conditioned discovery objectives (e.g., favoring familiar versus unfamiliar content). 
Under this formulation, podcast recommendation is inherently sequential: recent listening events provide strong signals of short-term intent, which evolves as users discover new topics or transition across successive sessions.

A natural baseline for sequential prediction is a transformer-based recommender such as SASRec~\cite{kang2018self} or TIGER~\cite{rajput2024recommender}. However, our production setting imposes two additional requirements that motivate a language-grounded formulation. First, recommendation objectives are not monolithic: different surfaces and usages impose different notions of relevance. For example, recommendations shown on a discovery-oriented surface may prioritize novelty and exploration, while those presented mid-session may favor topical continuity, familiarity, or format constraints such as episode length. 
Second, effective discovery requires language-level understanding to surface topics and themes that align with a user’s interests, beyond what can be captured by interaction patterns alone

These requirements motivate a formulation in which recommendation objectives and user context are expressed directly in language.
We therefore formulate recommendation as instruction-conditioned sequence modeling over a discretized, semantically grounded catalog. To satisfy production serving constraints, we focus on compact open-weight LLM backbones (up to $\sim$1B parameters) that balance expressiveness with latency and cost efficiency. The next section describes the model architecture, conditioning signals, and training procedure.

\section{Methodology}

For each user $u$ with interaction history $\mathcal{H}_u = \{(e_1,t_1),\ldots,(e_n,t_n)\}$, where $e_i$ denotes an episode and $t_i$ its timestamp, our goal is to generate a ranked list of $k$ candidate episodes from \emph{non-habitual} shows that the user is likely to listen next. We operationalize this objective as conditional generation:
\[
p(\text{SID}(e)\mid \text{prompt}(u,\mathcal{H}_u,\text{instruction})),
\]
where the model generates one or more SID sequences  conditioned on user context, recent listening history, and a task instruction. These sequences are then mapped back to episode identifiers and used as candidates for downstream ranking.

The model consumes four types of inputs: 
(i) recent listening context, serialized as a time-ordered sequence of SIDs; 
(ii) lightweight user context expressed as natural-language tokens (e.g., locale or country, high-level affinity topics); 
(iii) long-term user preferences represented by a dense collaborative-filtering user embedding, injected as a soft prompt~\cite{lester2021power, mu2023learning, huber2025embedding}; and
(iv) a natural-language instruction specifying the recommendation objective (e.g., non-habitual familiar versus non-habitual unfamiliar). 
The model generates SID sequences conditioned on these inputs, which are decoded via a lookup table into episode identifiers and returned as a ranked candidate set.

Figure~\ref{fig:architecture} shows that GLIDE is composed of two training stages to decouple catalog grounding from task-specific personalization. First, the model is trained to translate between SID sequences and episode text descriptors, making SIDs semantically meaningful to the model. Then, the grounded model is trained for instruction following tasks: given a prompt containing recent-history SIDs, long- and short-term user context, the model generates SID sequences corresponding to \emph{non-habitual}  episode recommendations.

\subsection{Semantic IDs and User Representation}
\label{sec:semantic_ids}
To enable generation over large, dynamic catalog, we represent each episode using a short sequence of discrete SIDs~\cite{rajput2024recommender, penha2025semantic}, obtained by quantizing a continuous episode content embedding (computed from the episode's title and description text) into a fixed-length sequence of indices. This discretization yields a compact vocabulary whose size scales with the quantizer configuration, rather than with the number of catalog items, making SIDs more efficient.

Episode embeddings are produced by encoding titles and descriptions with a proprietary text encoder fine-tuned on podcast-specific data. The encoder follows the architecture of multilingual BGE-M3~\cite{bge-m3}.

Although prior work has explored constructing SIDs from collaborative signals (CF) (e.g., ~\cite{2024_TokenRecLearningTokenizeIDLLMbaseda,2024_AdaptingLargeLanguageModelsIntegrating,penha2025semantic}) we focus on content-based SIDs derived from episode titles and descriptions for two reasons. First, \emph{cold-start coverage}: new episodes are published continuously and must be recommendable immediately, whereas CF embeddings require sufficient interaction volume before a stable representation emerges. Second, \emph{stability}: because content-based SIDs are derived from episode embeddings, updates to the embedding space propagates to SID assignments, while CF embeddings evolve with user behavior, making the episode-to-SID mapping non-stationary and introduce train--serve mismatch (i.e., different item representations at training and inference time). Content-based embeddings remain stable unless episode metadata itself is revised. 

To construct discrete SIDs from continuous episode embeddings, we require a quantization method that is scalable, stable over time, and compatible with auto-regressive generation.

\subsubsection{Residual K-Means quantization}
We construct SIDs using a residual quantizer, specifically Residual K-Means (R-KMeans)~\cite{penha2025semantic}, which balances simplicity with stable performance~\cite{ju2025generative}. R-KMeans encodes an episode embedding $x_e \in \mathbb{R}^d$ as a length-$M$ sequence of discrete cluster indices  by iteratively quantizing the residual error. At each level $m \in \{1,\ldots,M\}$, the algorithm selects the nearest centroid from a level-specific set of $K$ centroids $\{\mu_{m,1},\ldots,\mu_{m,K}\}$ and updates the residual:
\begin{align}
r_0 &= x_e, \\
c_m &= \arg\min_{j \in [1..K]} \left\lVert r_{m-1} - \mu_{m,j} \right\rVert_2, \\
r_m &= r_{m-1} - \mu_{m,c_m}.
\end{align}
The resulting SID is the tuple $(c_1,\ldots,c_M)$, which we treat as a sequence of discrete tokens.

In our implementation, the quantization level $m$ is represented implicitly via a distinct token namespace per level: for each level, we introduce $K$ tokens of the form $\langle\texttt{SID}_m{=}j\rangle$ for $j \in [1..K]$. With $K{=}256$ and $M{=}4$, this yields $M\times K = 1024$ SID tokens to the LLM vocabulary, and each episode is represented by exactly four tokens (one per level). For example, an episode may be encoded as $\langle\texttt{SID}_1{=}13\rangle\,\langle\texttt{SID}_2{=}65\rangle\,\langle\texttt{SID}_3{=}188\rangle\,\langle\texttt{SID}_4{=}7\rangle$.

As shown in prior work~\cite{singh2024better}, using multiple quantization levels can induce a coarse-to-fine semantic structure (e.g., \emph{News} $\rightarrow$ \emph{Sports-News}), which provides a useful inductive bias for auto-regressive generation.

\subsubsection{Soft Prompt User Personalization}
\label{sec:uservec}

Achieving user-level personalization with an LLM is challenging because short-term context alone (e.g., recent listening history) is often insufficient to capture stable, long-term preferences. A natural approach is to serialize user information in text, describing interests, topics, and historical behavior~\cite{huber2025embedding}, which results in long context. 

Instead, we inject long-term preference signals into the LLM via dense user embeddings from an existing production personalization model. Following prior work on embedding-based conditioning~\cite{huber2025embedding}, given a user embedding $v_u$, we project it into the LLM hidden dimension $d_{\text{LLM}}$ using a learned projection. The resulting representation forms a single soft prompt token $\tilde{V}_u = \mathrm{Proj}(v_u) \in \mathbb{R}^{1d \times d_{\text{LLM}}}$, computed via a two-layer MLP and inserted at the beginning of the user context, immediately following the system instruction. This soft prompt allows the LLM to attend to high-dimensional user state alongside natural-language tokens and SIDs, preserving a compact context window while maintaining rich collaborative signals.

\subsection{Semantic Grounding}
\label{sec:grounding}


Newly added SID tokens lack semantic grounding for the LLM, even when their embeddings are initialized in the base model’s embedding space~\cite{hewitt2021initializing}. Moreover, na\"ively adding new vocabulary items and fine-tuning the full model can destabilize pre-trained representations, leading to degraded language capabilities such as representation collapse or catastrophic forgetting~\cite{mundra2024empirical, biderman2024lora}. 

We therefore ground and align SID tokens to the base LLM through continued training on a bidirectional translation objective. For each episode, we pair its SID tuple $s_e$ with a textual descriptor $t_e$ derived from metadata (e.g., a summary or topic labels), and optimize both directions: generating $t_e$ from $s_e$ and generating $s_e$ from $t_e$. This bidirectional grounding objective encourages the model to treat SID tokens as semantically meaningful vocabulary items rather than arbitrary codes, enabling reasoning over catalog identifiers within the same representational space as natural language.

To ensure stable alignment, we adopt a two-stage adaptation strategy. First, we freeze the Transformer backbone and optimize only the newly initialized SID token embeddings, establishing a robust semantic projection for the catalog without perturbing the pre-trained model. Second, we freeze the entire model, including the embeddings, and adapt the Transformer blocks using LoRA~\cite{hu2022lora}. Compared to full-parameter fine-tuning, this parameter-efficient approach regularizes adaptation and helps preserve general language and world knowledge while improving catalog-specific grounding.


We do not ground the user soft prompting to maintain this stage free from collaborative signals, which are incorporated in the Instruction Tuning stage.


\subsection{Instruction Tuning}
\label{sec:sft}

\begin{figure}[t]
    \centering
    \small
    \begin{tcolorbox}[colback=gray!5!white,colframe=black!70,boxrule=0.8pt,arc=3pt,width=0.9\linewidth]
        \textbf{[System Instruction]} \\
        \textit{Define the role (Recommender) and output format (Semantic IDs).}
        
        \vspace{0.2cm}
        \textbf{[User Context]} \\
        $\texttt{<Soft Prompt Vector>} \oplus \texttt{<Textual Metadata>}$
        
        \vspace{0.2cm}
        \textbf{[Interaction History]} \\
        \texttt{<Sequence of Recent Episode SIDs>}
        
        \vspace{0.2cm}
        \textbf{[Task Instruction]} \\
        \textit{"Recommend a \textbf{\{\{ familiarity\_mode \}\}} episode aligned with user interests."}
    \end{tcolorbox}
    \caption{Structure of the GLIDE prompt. The model is conditioned on a hybrid of continuous soft prompts and discrete tokens.}
    \label{fig:prompt_template}
\end{figure}

We employ instruction tuning to adapt the semantically grounded model to the downstream task of episode recommendation. We formulate the input prompt as a unified, natural language directive that aggregates three signal modalities into a single context window, as illustrated in Figure~\ref{fig:prompt_template}: 
\begin{enumerate} 
\item \textbf{User Representation:} A hybrid context comprising the projected dense user embedding (soft prompt) and high-level textual metadata (e.g., locale, affinity topics). 
\item \textbf{Interaction History:} A time-ordered sequence of SIDs representing the user's recent listening sessions. 
\item \textbf{Task Instruction:} A final natural language directive containing a dynamic control token to specify the recommendation goal. 
\end{enumerate}

Crucially, we introduce and optimize the learned projection jointly with the LLM parameters during this stage. This allows the model to learn the mapping between the collaborative filtering space—which was excluded during the Semantic Grounding phase—and the LLM's internal representation.

This design enables a single model to support multiple discovery objectives. By simply swapping the control token at inference time (e.g., from \textit{“familiar”} to \textit{“unfamiliar”}), we can explicitly shift the generation distribution toward either previously known shows or entirely novel content without retraining.

\subsubsection{Multi-task learning for controllable discovery.}
\label{sec:multi-task}
Training on all non-habitual streams with a single objective increases the tendency of the model toward non-habitual but familiar shows, which are more commonly observed in interaction logs than unfamiliar content. To address this, we employ multi-task supervised fine-tuning to explicitly distinguish between these two discovery horizons.

We operationalize this using the prompt structure shown in Figure~\ref{fig:prompt_template}. For each training example, we determine the target's relationship to the user's history and populate the control token accordingly: \begin{itemize} \item \textbf{Familiar Mode:} If the target episode belongs to a show the user has previously listened to (but is not currently habitual), we set the control token to \texttt{familiar}. \item \textbf{Unfamiliar Mode:} If the target episode belongs to a show the user has never listened to, we set the control token to \texttt{unfamiliar}. \end{itemize}

This formulation enables the model to disentangle the two objectives within a single set of parameters. By learning to associate the familiarity mode token with the corresponding retrieval distribution, the model gains the ability to be explicitly conditioned at inference time, allowing the model to request "pure discovery" or "familiar discovery" depending on the surface context.

We mitigate skewed prediction distributions that arise from 
exposure effects and popularity bias using the following strategies:
\begin{itemize}
  \item \textbf{Cross-surface sampling:} 
  We draw training examples from multiple surfaces to reduce feedback loops in which exposure from one surface disproportionately reinforces its own recommendations over time.
  \item \textbf{Exploration upweighting:} 
  We assign higher sampling weight to streams originating from randomized or exploration-driven placements, as these interactions are less shaped by prior recommender exposure and therefore provide cleaner signals of user preference.
  \item \textbf{Popularity capping:} 
  We limit the number of training examples contributed by any single episode, preventing highly popular items from dominating the training loss and reinforcing popularity bias.
\end{itemize}

Together, these training choices balance relevance, exploration, and robustness, preparing the model for deployment under real-world serving constraints. 



\subsection{Model Serving}
\label{sec:system}

Deploying GLIDE in production requires addressing system-level challenges in serving, scalability, and reliability, in addition to model accuracy.

\subsubsection{Collision resolution.}
\label{sec:collision}
Quantizing embeddings into SIDs can map multiple episodes to the same SID sequence, resulting in collisions~\cite{rajput2024recommender, singh2024better}. Internal analyses indicate that such collisions typically arise from near-duplicate content embeddings, often caused by episodes with short or repetitive descriptions, or by multiple episodes within the same show sharing highly similar metadata.

At inference time, we resolve collisions using a deterministic popularity-based tie-breaker.  For each episode sharing the same SID, we select the most popular episode that satisfies eligibility constraints (e.g., playable, available in the user’s locale, and absent from the current candidate set).
Since collisions generally occur among semantically near-indistinguishable items under the embedding model, this strategy resolves ambiguity without additional model inference, using  popularity as a pragmatic proxy for quality within each group. The tie-breaker is updated daily.

\subsubsection{Deployment}
Deploying an LLM-based recommender in production requires meeting strict constraints on latency, cost efficiency, and operational reliability. We deploy GLIDE as an online inference service integrated into existing recommendation pipelines. Requests are event-triggered (e.g., session start or user-state updates), which allows results to be pre-computed and cached when appropriate while still meeting downstream service-level objectives.

At inference time, the service assembles a compact model input from (i) recent interaction history represented as SIDs, (ii) lightweight user context, and (iii) a dense long-term user representation injected as a soft prompt. The model generates candidate SID sequences, which are mapped back to concrete episode identifiers and passed to downstream ranking components.

\subsubsection{Soft Prompts Support}
GLIDE conditions the SID generation on dense user embeddings via soft prompts (\Cref{sec:uservec}). We implement per-request conditioning by passing the projected user embedding alongside the tokenized prompt and prepend it at a fixed placeholder position during input preparation. This preserves a stable prompt structure between training and serving while keeping the inference engine interface unchanged. The added overhead is constant per request; the dominant runtime cost remains autoregressive decoding.

\subsubsection{Beam Search Optimization}
\label{sec:beam}
Consistent with prior work on catalog-grounded generative retrieval~\cite{2025_PLUMAdaptingPretrainedLanguageModels, 2024_AdaptingLargeLanguageModelsIntegrating, 2024_TokenRecLearningTokenizeIDLLMbaseda}, we use Beam Search with 30 beams to generate 30 candidates per request. Each generated SID sequence is mapped to episode identifiers via a lookup table and passed to downstream ranking stages.

In our initial deployment, wide-beam decoding exposed two coupled bottlenecks. First, CPU-side request orchestration increased latency under load. Second, accelerators were under-utilized due to the many small, sequential decoding steps induced by beam expansion. Rather than modifying the inference engine, we mitigated these issues with standard serving optimizations, including scaling the request-orchestration layer and tuning the serving configuration to better match the utilization profile of wide-beam decoding. Under the same latency objectives, these changes improved throughput by up to 8$\times$, enabling an increase in beam width from 14 to 30 while maintaining production latency targets.

\section{Experiments and Results}
We evaluate the performance of the proposed GLIDE framework through both offline and online experiments. Our generative recommender model was initialized with a Llama 3.2 1B–based backbone model \cite{dubey2024llama} before adaptation.

\subsection{Evaluation Framework}
To achieve a robust evaluation, we rely on a framework composed of three complementary  components with different strengths, costs and levels of fidelity: traditional retrieval metrics computed from interaction logs, human evaluation, and LLM-based judges.

\paragraph{Retrieval Metrics.}
We compute Recall@30, HitRate@30 and NDCG@30 on held-out non-habitual streams. These metrics are reported separately for familiar and unfamiliar content segments (as defined in Section~\ref{sec: problem_design}). This allows us to distinguish improvements driven by better modeling of latent preferences (familiar) from gains due to genuine expansion of user taste (unfamiliar).

\paragraph{Human Evaluation.}
We complement offline metrics with internal human evaluations performed with Spotify employees. In each evaluation round, annotators review sampled top-K recommendation lists and rate recommendations along qualitative dimensions including interest alignment (topic, host or guest, and style or format), recency or freshness, diversity, and familiarity. Annotators also provide brief free-text comments to capture errors and subjective preferences. These evaluations reveal failure modes not captured by retrieval metrics, such as locale or language mismatches, and recommendations that are stale or already consumed. Insights from human evaluation are used to guide model iteration and to calibrate LLM-based judge assessments.

\paragraph{LLM-judges.}
To scale qualitative evaluation beyond what human evaluation sessions can support, we leverage LLM-based judges to assess recommendation quality. Previous work has shown that LLM judges align well with human feedback in the podcast recommendation problem~\cite{fabbri2025evaluating}. 

The LLM judge takes as input a user profile summarizing top shows, episodes, and topics, together with the recommended episode metadata and a transcript excerpt. We prompt the judge to assess whether the episode is interest-aligned with the user by decomposing alignment into pointwise dimensions--\textit{topic}, \textit{host \& guest}, \textit{style \& format}, \textit{tone \& genre}--along with an overall \textit{interest-aligned} score. We also consider listwise dimensions to evaluate whether recommendation lists are \textit{interest-diverse}, covering multiple interests, and \textit{interest-representative}, capturing the breadth of the user profile. This approach complements offline retrieval metrics with scalable, fine-grained assessments of recommendation quality. 

\subsection{Offline Results}
We first present offline results that quantify the impact of different model variants on retrieval quality for non-habitual content.
Datasets are constructed by sampling user interaction sequences from a 7-day sliding window of streaming logs with a user-based evaluation split.

\subsubsection{Episode Recommendation}
We evaluate the contribution of different conditioning signals in the proposed language-grounded recommendation framework by benchmarking GLIDE against two LLM-based variants. All models operate over the same semantically grounded LLM backbone (Section~\ref{sec:grounding}), differing only in their input conditioning:
\begin{itemize}
    \item \textbf{SID-only}: A reference baseline that models recommendation purely as sequence generation over SIDs, without additional semantic or personalization signals. This setting is similar to TIGER~\cite{rajput2024recommender};
    \item \textbf{SID + Text}: An ablation that augments the SID sequence with textual episode information, but excludes the user soft prompt.
\end{itemize}

As summarized in Table~\ref{tab:offline}, incorporating textual information (\textit{SID + Text}) yields substantial gains over SID-only baseline, suggesting that explicit text enables the model to better leverage its pre-trained semantic knowledge. GLIDE achieves the strongest overall performance, with a +29.9\% improvement in Recall@30 and a +31.2\% improvement in NDCG@30. These results indicate that combining semantic grounding with user-level personalization improves both candidate retrieval and ranking quality.

The benefits of GLIDE are most pronounced in the \textit{Non-Habitual Unfamiliar} segment, which is central to the discovery objective. In this setting, GLIDE significantly outperforms the text-conditioned variant (+35.4\% vs. +14.7\% in NDCG@30), indicating that the full training recipe is particularly effective at recommending content outside a listener's established routine, where interaction-based signals are weakest.

\begin{table}[t]
  \centering
  \caption{Offline evaluation. Relative improvement (\%) over the baseline across content familiarity segments.}
  \label{tab:offline}
  \begin{tabularx}{0.48\textwidth}{@{}Xrr@{}}
   \toprule
    \textbf{Method} 

      &\textbf{Recall@30} 
      & \textbf{NDCG@30} \\
    \midrule
     \multicolumn{3}{c}{\cellcolor{gray!15}\textit{Overall}} \\
     LLM (SID only) (Baseline) & - & - \\
     LLM (SID + Text) & +25.0\% & +28.2\%\\
     \rowcolor{blue!5}
     GLIDE (Ours) & +29.9\% & +31.2\% \\
    \midrule
\multicolumn{3}{c}{\cellcolor{gray!15}\textit{Non-Habitual but Familiar}} \\
     LLM (SID only) (Baseline) & - & - \\
     LLM (SID + Text) & +25.0\% & +30.5\%\\
     \rowcolor{blue!5}
     GLIDE (Ours) & +28.7\% & +28.7\% \\
    \midrule
    \multicolumn{3}{c}{\cellcolor{gray!15}\textit{Non-Habitual and Unfamiliar}} \\
     LLM (SID only) (Baseline) & - & - \\
     LLM (SID + Text) & +18.1\% & +14.7\%\\
     \rowcolor{blue!5}
     GLIDE (Ours) & +29.5\% & +35.4\% \\
    \bottomrule
  \end{tabularx}
\end{table}

\subsubsection{What is the best quantizer for SIDs?}
We compare three approaches to constructing SIDs from episode content embeddings: RQ-VAE, R-LFQ, and R-KMeans. All three methods generate a short sequence of discrete codes for each episode by progressively quantizing the embedding, but they differ substantially in complexity and training requirements.

RQ-VAE~\cite{lee2022autoregressive} is a neural quantization approach that learns discrete codes through end-to-end training with a learned codebook. It has been widely used in prior work on generative retrieval~\cite{rajput2024recommender}, but is known to be difficult to train reliably, often requiring careful tuning and additional stabilization techniques to prevent under-utilization of codes~\cite{huh2023straightening,lancucki2020robust}. 
R-LFQ builds on Lookup-Free Quantization (LFQ)~\cite{yu2024language}, which replaces explicit codebooks with simple binary decisions in the latent space. We extend this method to a residual setting by applying it iteratively. While this approach avoids maintaining large codebooks, it still relies on end-to-end training with regularization to ensure effective use of codes, adding training complexity.

Table~\ref{tab:semid_results} compares offline retrieval performance and semantic consistency across methods. R-KMeans achieves a 9.52\% relative improvement in Hit-Rate@30 over RQ-VAE, and substantially outperforms R-LFQ (+4.76\%). This advantage is accompanied by higher semantic consistency: episodes assigned the same SID under R-KMeans have a higher average cosine similarity in the original embedding space (0.856) than those under RQ-VAE (0.657). This property is particularly important in our setting, as higher within-ID semantic consistency ensures that collision resolution (Section~\ref{sec:collision}) surfaces meaningful alternatives when the target episode is filtered.

\begin{table}[t]
\centering
\caption{Ablation across SID quantization methods. R-KMeans achieves better accuracy and higher embedding cosine similarity within each group (sharing the same SID) where there is at least one collision.}
\label{tab:semid_results}
\begin{tabularx}{0.48\textwidth}{@{}Xrr@{}}
\toprule
\textbf{SemID Space} & \textbf{HR@30} & \textbf{Intra-Bucket Similarity} \\ \midrule
RQ-VAE (Baseline)    & --                   & 0.657                            \\
R-LFQ                & +4.76\%              & 0.713                            \\
\rowcolor{blue!5}
R-KMeans             & +9.52\%     & 0.856                   \\ \bottomrule
\end{tabularx}
\end{table}

Deep quantizers such as RQ-VAE and R-LFQ offer appealing theoretical properties: their end-to-end training makes it possible to directly encode desired behavior into the quantization process, for example by encouraging similar items to share codes~\cite{2025_PLUMAdaptingPretrainedLanguageModels} or by promoting balanced code usage. However, these methods also introduce substantial engineering complexity. In practice, training can be unstable, with issues such as codebook collapse, sensitivity to initialization, and dependence on carefully tuned auxiliary losses, making them difficult to deploy reliably in production systems~\cite{dhariwal2020jukebox,huh2023straightening}. 
Given the strong empirical performance and operational simplicity of R-KMeans, we adopt it as our production quantizer. 

\subsubsection{Impact of Controllable Multi-Task Learning.}
We validate the multi-task strategy (Section \ref{sec:multi-task}) by comparing GLIDE Multi-Task against a GLIDE Single-Task variant. We define the single-task model as one trained on the aggregate dataset without control tokens, effectively treating all non-habitual listening as a single, task. Without explicit signals, this single-task variant models the marginal distribution over mixed intents, resulting in predictions that collapse toward the dominant mode of the training data.

By contrast, GLIDE Multi-Task disentangles these objectives. When conditioning the generation on the unfamiliar token, we observe a +11.8\% relative improvement in Recall@30 for Non-Habitual Unfamiliar items compared to the single-task baseline. Similarly, prompting with the familiar mode results in a +4.9\% lift for Non-Habitual Familiar content. This confirms that conditioned multi-task learning effectively specializes the output distribution for specific discovery horizons while simultaneously improving the performance on non-habitual familiar content.

\subsubsection{Ablation: Semantic Grounding.} 
We evaluate the impact of the Semantic Grounding phase (Section~\ref{sec:grounding}) by comparing two model variants that both include user-level personalization via the soft prompt. In the first variant, the model is instruction-tuned directly for recommendation without first grounding the SID tokens. In the second one, we use the full GLIDE pipeline, in which SIDs are first aligned with text through a bidirectional translation objective before instruction tuning.

Semantic Grounding yields consistent improvements across metrics, yielding a +8.34\% relative lift in Recall@5. These results suggest that connecting SID tokens with their textual meaning provides a more informative starting point for the recommendation task. Intuitively, grounding helps the model treat SIDs as meaningful representations of content, rather than as arbitrary symbols.

\subsubsection{Ablation: Beam Search} As mentioned in \Cref{sec:beam}, we rely on beam search at inference time, but its necessity is not obvious given the added computational cost. We therefore evaluate its impact by comparing beam search against simpler sampling-based decoding strategies, such as greedy decoding and sampling with temperature or top-$p$ (Table \ref{tab:beam_search_degradation}). 

Replacing beam search with greedy decoding results in a 27.07\% relative drop in Recall@30, indicating a substantial loss in retrieval quality. To understand the source of this loss, we analyze the Prefix Ceiling, which measures the maximum achievable recall when considering only the first few tokens of the generated SID sequence. We see that most relevant candidates are captured within the first two tokens, with only a 12.53\% drop in recall. However, performance degrades sharply as longer sequences are required.

This pattern suggests that sampling-based decoding is often sufficient to identify the coarse semantic region of the target episode, but struggles to select the correct fine-grained identifier within that region. As a result, generation frequently collapses to low-quality or invalid SIDs. These findings show that beam search is necessary to reliably navigate the fine-grained structure of SIDs and recover high-quality recommendations.

\begin{table}[t]
\centering
\caption{Impact of different decoding strategies. We report the relative degradation when moving from Beam Search to Greedy Decoding. To analyze the source of the degradation, we provide the Prefix Ceiling (the maximum achievable Recall given the first $N$ tokens of the generated SIDs).}
\label{tab:beam_search_degradation}
\begin{tabular}{@{}llr@{}}
\toprule
\textbf{Decoding Strategy} & \textbf{Scope} & \textbf{$\Delta$ Recall@30} \\ \midrule
\rowcolor{blue!5}
Beam Search & Full Sequence (4 tokens)  & --        \\
Sampling    & Full Sequence (4 tokens)  & -27.07\%  \\
Sampling    & Prefix Ceiling (3 tokens) & -26.17\%  \\
Sampling    & Prefix Ceiling (2 tokens) & -12.53\%  \\ 
\bottomrule
\end{tabular}
\end{table}

\subsubsection{LLM-judge Results}

\label{sec:judge_validation}
Using LLMs as evaluators has recently emerged as a scalable alternative to human annotation~\cite{zheng2023judging}. In information retrieval, prior work has shown that LLMs such as GPT-4 can predict document relevance with near-human accuracy~\cite{thomas2024large}. Recommendation evaluation, however, introduces additional challenges, since user preferences must be inferred from past behavior, and recommendations do not come with explicit queries to anchor judgments.

Recent work on profile-aware evaluation~\cite{fabbri2025evaluating} addresses these challenges by representing user interests as natural-language profiles, enabling LLMs to assess whether recommended items align with a user's preferences. We adopt this approach and continuously validate the LLM-judge by comparing its assessments with feedback collected from internal human evaluations. We observe that judge scores correlate positively with Recall@30, while also capturing qualitative aspects of recommendation quality that are not reflected in retrieval metrics alone.

Importantly, in one evaluation round, the LLM-judge and human evaluators agreed on a preferred model variant, while offline recall metrics favored a different one. Further analysis showed that the recall-favored model exhibited stronger popularity bias: it achieved higher hit rates by recommending more popular episodes, which were judged as less interest-aligned. This case highlights the complementary role of LLM-judges in identifying quality differences that offline metrics can miss.

\begin{figure}[t]
\centering
\includegraphics[width=0.8\columnwidth]{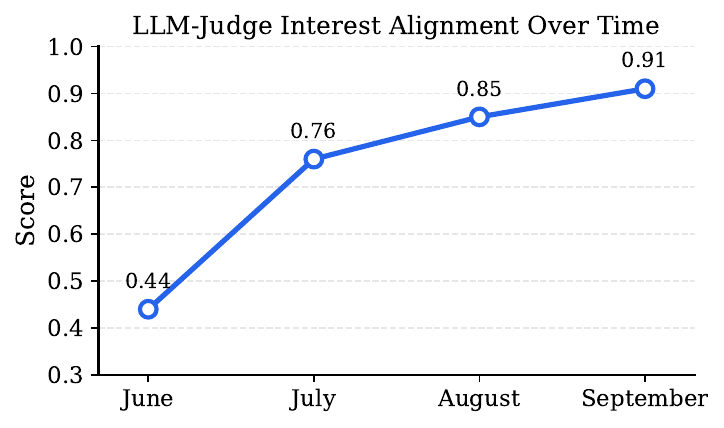}
\caption{Interest alignment scores from the LLM-judge across multiple model development iterations.}
\label{fig:interest_aligned}
\end{figure}

Throughout development, we tracked LLM-judge scores alongside traditional metrics and observed steady improvements in interest alignment as we iterated on the training recipe and data quality (Figure~\ref{fig:interest_aligned}). For a more detailed analysis of how LLM-judge rankings relate to traditional retrieval metrics, and how they help mitigate exposure and popularity biases, we refer the reader to~\cite{penha2025llmjudges}.\\

Taken together, these offline results show that semantic grounding, user-level personalization, and careful decoding are all critical to improving retrieval quality for non-habitual content, motivating validation in an online setting.

\subsection{Online A/B Test}
\label{sec:online_ab}

We conducted a 21-day randomized user-level online experiment on podcast listeners in English-speaking markets who where active in the 30 days prior to evaluate the effect of generative recommendations on discovery for engaged users. We compared a Control condition, using the existing production recommendation system, against a single treatment condition. In the treatment, candidates generated by GLIDE were added as an additional candidate retrieval source to the existing candidate pool and then scored and ranked by the standard downstream episode ranker alongside candidates from other sources.

We report results on the Home surface of Spotify, with each cell covering approximately 20M impressions. Adding GLIDE as a candidate source led to statistically significant improvements in non-habitual consumption and new show discovery, while preserving guardrails and with no regressions in overall engagement or user satisfaction. Specifically, in the treatment, non-habitual streams per user increased by 5.4\% ($\alpha <0.01$) and new-show non-habitual streams increased by 14.3\% ($\alpha <0.01$). GLIDE candidates accounted for $\sim34\%$ of recommendations in the treatment group, supporting the attribution of the observed lift to the added candidate source. 
The system met production cost and latency constraints, remaining within the allocated serving budget for candidate generation. 

These results demonstrate that language-grounded, generative recommendation can drive meaningful discovery gains at scale while satisfying real-world deployment requirements.


\section{Conclusion}

We presented a production-ready, LLM-based generative recommender for episode-level podcast discovery, grounded in SIDs and designed for deployment at scale. The proposed system combines language-conditioned generation with catalog grounding and user-level personalization, enabling controllable, session-aware recommendations over a large and dynamic content catalog.

Through extensive offline experiments, we showed that semantic grounding, user personalization, and careful decoding choices each contribute meaningfully to retrieval quality for non-habitual content. These gains translated to real-world impact in a 21-day online A/B test on Spotify home surface, where the system increased non-habitual podcast streams by +5.4\% and new-show non-habitual streaming by +14.3\%, with no regressions in engagement or user satisfaction and within production cost and latency constraints.

Beyond model performance, our results highlight the importance of evaluation design. We showed that offline retrieval metrics alone are insufficient to capture discovery-oriented success, and that combining offline metrics with human judgments and LLM-based evaluation provides a more complete picture of recommendation quality.

Future work includes deeper analysis of how grounded language models leverage pre-trained world knowledge in recommendation settings, extending language-conditioned control to additional objectives and surfaces, and further improving evaluation methods for long-term recommendations.

\begin{acks}
The authors would like to thank Oskar Stål for his technical leadership and invaluable contributions to the project.
\end{acks}

\bibliographystyle{acm}
\bibliography{main}


\end{document}